\documentclass{jaa}
\usepackage{natbib}
\usepackage{xcolor}
\definecolor{xlinkcolor}{cmyk}{1,1,0,0}

 \usepackage[bookmarks=false,         % show bookmarks bar?
     pdfnewwindow=true,      % links in new window
     colorlinks=true,    % false: boxed links; true: colored links
     linkcolor=xlinkcolor,     % color of internal links
     citecolor=xlinkcolor,     % color of links to bibliography
     filecolor=xlinkcolor,  % color of file links
     urlcolor=xlinkcolor,      % color of external links
final=true
 ]{hyperref}

\usepackage{graphicx}
\usepackage{lastpage}
\usepackage{multicol}
\usepackage{aas_macros}
\usepackage{subcaption}
\usepackage{rotating}
\usepackage{tablefootnote}
\usepackage{multirow}
\usepackage{subfloat}
\usepackage{subcaption}
\usepackage{hyperref}
\usepackage{soul}

\newcommand{\degr}{\ensuremath{^\circ}}
\newcommand{\asat}{{\em AstroSat}}
\newcommand{\fermi}{{\em Fermi}}

\newcommand{\sw}[1]{\texttt{#1}}
\newcommand{\geant}{\sw{GEANT4}}
\newcommand{\gray}{\ensuremath{\gamma}\textendash ray}

\hypersetup{colorlinks=true,allcolors={cyan}}
\newcommand{\rough}[1]{#1}

\newcommand{\xs}{\texttt{xspec}}

\begin{document}\sloppy

\title{Revisiting the Earth's atmospheric scattering of X-ray/\gray s and its effect on space observation: Implication for GRB spectral analysis}

\author{
\href{https://orcid.org/0000-0003-2932-3666}{Sourav Palit}\textsuperscript{1*},
\href{https://orcid.org/0000-0002-8935-9882}{Akash Anumarlapudi}\textsuperscript{1,2},
\and
\href{https://orcid.org/0000-0002-6112-7609}{Varun Bhalerao}\textsuperscript{1}
}

\affilOne{\textsuperscript{1} Department of Physics, Indian Institute of Technology Bombay, Mumbai, India\\}
\affilTwo{\textsuperscript{2} Department of Physics, University of Wisconsin-Milwaukee, Milwaukee, Wisconsin, USA\\}

\twocolumn[{

\maketitle

\corres{souravspace@iitb.ac.in}

%\msinfo{1 January 2015}{1 January 2015}

\begin{abstract}
A considerable fraction of incident high energy photons from astrophysical transients such as Gamma Ray Bursts (GRBs) is Compton scattered by the Earth's atmosphere. These photons, sometimes referred to as the ``reflection component'', contribute to the signal detected by space-borne X-ray/\gray\ instruments. The effectiveness and reliability of source parameters such as position, flux, spectra and polarization, inferred by these instruments are therefore highly dependent on the accurate estimation of this scattered component. Current missions use dedicated response matrices to account for these effects. However, these databases are not readily adaptable for other missions, including many upcoming transient search and gravitational wave  high-energy Electromagnetic counter part detectors. Furthermore, possible systematic effects in these complex simulations have not been thoroughly examined and verified in literature. We are in the process of investigation of the effect with a detailed Monte Carlo simulations in \geant\ for a Low Earth Orbit (LEO) X-ray detector. Here, we discuss the outcome of our simulation in form of Atmospheric Response Matrix (ARM) and its implications of any systematic errors in the determination of source spectral characteristics. We intend to apply our results in data processing and analysis for \asat-CZTI observation of such sources in near future. Our simulation output and source codes will be made publicly available for use by the large number of upcoming high energy transient missions, as well as for scrutiny and systematic comparisons with other missions.
\end{abstract}

\keywords{GRB, Atmospheric scattering, X-ray, GEANT4}
}]

\doinum{12.3456/s78910-011-012-3}
\artcitid{\#\#\#\#}
\volnum{000}
\year{0000}
\pgrange{1--}
\setcounter{page}{1}
\lp{\pageref{LastPage}}

\section{Introduction}\label{sec:intro}

X-ray/\gray\ astrophysical observations are primarily conducted from space-based platforms as the Earth's atmosphere does not transmit these high energy photons. While most photons are absorbed, a fraction of the X-ray/\gray\ photons from any extra-terrestrial sources undergoes Compton scattering in the upper layers of the atmosphere. In the Earth's atmosphere, which has a mass column density of $\sim 10^3~\mathrm{g~cm}^{-2}$ at sea level, some of the scattered photons travel upwards having an appearence of reflection of X-ray/\gray\ by the atmosphere. Any space borne measurement of those high energy photons from astrophysical sources will be affected by simultaneous detection of the reflected photons.

The topic of interaction of extra-terrestrial high energy photons with the
atmosphere is not new. Various studies have been performed to determine the
spectrum of reflected Cosmic X-ray Background (CXB)\footnote{Alternately termed
as Cosmic Diffused Gamma ray Background (CDGRB).} photons from the Earth's
atmosphere and their effects on observations by various space-based high-energy
instruments \citep[e.g.][]{churazov2008}. Such works are often done in
conjugation with studies determining the X-ray/\gray\ spectrum arising
from the interaction of Cosmic ray particles \citep[e.g.][]{sazonov2017} with
the molecules in Earth's atmosphere. Together these two components constitute
the X-ray/\gray\ albedo contribution and
add up to the total background noise in high energy space detectors.

Apart from such studies in the context of Earth's atmosphere, the main
motivation of research related to the Compton reflection of high energy photons
has been of purely astrophysical nature. This includes mainly the study of the
scattering during the passage and interaction of cool electrons in relatively
cooled plasma, such as the corona above a relatively cool, optically thick
accretion disk \citep[e.g.][]{Bisnovatyi-Kogan_77}. In such studies one usually
calculates a Green's function describing the reflection of monochromatic
radiation \citep{White_1998}, and convolves the incident spectrum with this
function
to calculate the reflected spectrum.

From the studies of Earth's atmospheric scattering of high energy photons it is
estimated that a moderate fraction of the incoming X-ray/\gray\ photons is
reflected or back-scattered in the upper atmospheric layers of the Earth. Some
studies estimate that in the $\sim$ 30 -- 300~keV energy range (hereafter
referred to as ``intermediate energy") of X-rays/\gray s, as many as
$\sim30$\% of the incident photons can be scattered back \citep[see for
instance][]{churazov2008}.
The scattered fraction decreases outside this energy range, but remains non-negligible 
over a much broader X-ray/\gray\ band.This suggests that Low Earth Orbit (LEO) satellites 
observing energetic photons from Gamma Ray Bursts (GRBs) and other astrophysical transients 
detect significant flux reflected from the Earth's atmosphere as well. The magnitude and 
spectrum of the contribution should vary largely based on the relative angular position (viewing angle)
of the GRB  with respect to the instrument, as well as to the line joining the
 instrument to the geo-center, relative orientation of the instrument with the ground, and the source spectral 
properties. Whether this is considered as an interpretable addition to the source signal or merely as an 
addition to the background noise detected by the instrument like the CXB and albedo 
photons, any analysis of properties of the transients should correctly account for the contribution.

Till date, the contribution of Earth's reflection of X-rays and \gray s in
prompt GRB observation has mainly been studied from the standpoint of its
application in study of firstly, polarization in those wavelengths
\citep[e.g.][]{Willis2005} and secondly in the process of localization of
transient source \citep[e.g.][]{Pendleton_1999}.  The fact that the angular
distribution of Compton scattered flux from Earth's atmosphere should be
dependent on the direction of polarization of the incoming photon beam from a
distant GRB sources, has been exploited to find the polarization
characteristics from measurement of the reflected X-ray flux. There has been
rigorous investigation \citep{Pendleton_1999} for estimating the atmospheric
scattering contribution to accurately determine transient source location by
{\em CGRO}/BATSE and \fermi\ missions. They mainly concentrate on Monte Carlo
simulation studies and some observations
with solar flares.
%The atmospheric Compton scattering probability is high at the medium energy range (50 - 500 keV) of the prompt GRB spectrum and resulting  photons in the intermediate energy range should be readily detected by any space satellite along with the actual source photons, and hence bound to impart error in the estimation of the source spectrum itself.

%We have put some efforts in finding information on the method of mitigating the atmospheric scattering components in prompt GRB spectral processing in the mentioned energy range of interest.
We investigated how various missions have incorporated the Earth reflection
effect in their data processing.
We have been particularly interested in digging into the {\em CGRO}/BATSE and
\fermi/GBM spectral data processing and distribution, as these two have been
the main workhorses in providing the spectral observation of GRB prompt
emission from hard X-rays up to soft \gray\ part of the spectrum. In case of
CGRO/BATSE the atmospheric contribution is evaluated during burst localization
using EGS electromagnetic cascade Monte Carlo code with a spherical geometry of
concentric shells representing an atmosphere with exponential density.
Subsequent iterative $\chi^2$ minimization technique is followed
\citep{Pendleton_1999} to differentiate the scattering contribution from the
total observed spectra consisting of photons both directly detected from the
source and those scattered from atmosphere. 

The analyses presented by \cite{Pendleton_1999} show us the extent to which 
the atmospheric scattering of source photons can contaminate observation of transients.
In Figure 3 of the article demonstrating the distribution of number of GRBs detected 
by the detectors of different viewing angles (with respect to the burst) as function of 
the ratio of the detected atmospheric scattered flux to that of directly detected flux,  
it seems that for the detector with large viewing angles for a significant number of cases 
the detected scattered counts are well above the direct counts (by as much as an order). For 
moderate viewing angles the scattering rate is between 20\% and 50\% of that of the 
direct count rate. Figure 4 of the same article is also significant, depicting how the detected 
scattered to direct photon ratio can vary with zenith angle (i.e., the angle between the  
direction from instrument to the geo-center and the direction to the GRB), the viewing angle 
(w.r.t the burst) of detector and the orientation of the detector normal respective to the 
ground normal. The bottom-left plot showing the variation of the ratio with the source zenith 
angle for large viewing angle of the downward-facing detectors demonstrates the highest 
contribution of atmospheric scattering and its sharp dependence on photon incidence angle.

For \fermi/GBM, according to \cite{Meegan_2009}, the atmospheric scattering response 
estimation has been performed following the same method as described for BATSE in the GEANT4 
based GBM Response Simulation System \citep[GRESS;][]{Hoover_2005, Kippen_2007} and validated 
with BATSE simulation results. This is then used to form a source-specific total response 
function for a particular set of earth-spacecraft viewing conditions, to be calculated in the 
IODA DRMGEN software in stand alone mode. Whereas, \cite{Connaughton_2015} directly  use the 
atmospheric response database established for BATSE \citep{Pendleton_1999} in their study. 

The problem of unwanted contribution of atmospheric scattering of X-ray/\gray\
 is well addressed and steps have been followed to mitigate it, albeit, seemingly,
there has been no quantitative, comprehensive and generalized description on
how this should manifest in the estimation of the source characteristics, such
as photon spectrum and polarization. These studies have been highly focused on
specific instruments, and the final products are often deeply integrated with
the instrument response function. Instead, in this work we take a generalized
approach by concentrating on the Monte Carlo simulation of the interaction of
energetic astrophysical source photon at atmosphere, completely devoid of any
specific instrument interactions and characteristics. Thus, our goal is to calculate the
distribution of atmospheric Compton scattered photons for a plane--parallel
photon beam incident upon the Earth's atmosphere and how it should manifest on
the outcome of the spectral and polarization characteristics of the cumulative
photon detection.  Here we are presenting our findings only on
the spectral front and intend to address the same on polarization elsewhere.

The broader goal of this work is two-fold. The first is to replicate the 
work that has been performed for other missions, and apply it directly to analysis of sources in \asat-CZTI
\citep{2014SPIE.9144E..1SS,czti,Rao2017a}. Since its launch 5 years ago CZTI has
detected more than 400 GRBs \citep{sharma2020search}. CZTI has been
used for some GRB localization studies \citep{bkb+17}, and has also been very
successful in detection of polarization in them \citep[see][]{2019ApJ...884..123C}.
Inclusion of the process of evaluation and deduction of atmospheric scattered
contribution from prompt GRB spectra will be a significant step towards
accurate analysis of all the above characteristics.

The second goal of the work is to publicly release all source codes 
and simulation data in a format that will be easy to adapt to any other mission. 
This goal stems from the plurality of upcoming and proposed missions to study 
high energy transients and gravitational wave high-energy electromagnetic counterpart detection, 
such as large instruments like SVOM~\citep{Cordier2015}, GECAM~\citep{Zhang2019}; as 
well as cubesat class instruments like BurstCube~\citep{2017arXiv170809292R}, 
HERMES~\citep{2019NIMPA.936..199F}, BlackCAT~\citep{Chattopadhyay2018}, and 
Camelot~\citep{camelot}. There are also proposed missions, which are scientifically 
oriented towards hard X-ray/gammma ray polarimetry, such as 
POLAR-2 \citep{Hulsman:2020goi} on-board Chinese space station and a Large Area 
GRB Polarimeter (LEAP) \citep{McConnell_2017} on-board ISS. Rather than each of 
these missions developing their own atmospheric scattering databases, a single 
open database will serve the purpose well. It will also make it possible for 
comparisons between any other atmospheric scattering simulations, thus arriving 
at a quantitative estimate of systematic errors that may be present in the responses.

The organization of the paper is as follows. In Section~\ref{sec:simulation},
we discuss our simulation setup including detector geometry of Earth's
atmosphere and photon collecting detector, the physics of interaction and
photon generation and run execution process. In Section~\ref{sec:results} we
presents our simulation outcome for mono-energetic photons and a hypothetical
GRB spectrum, represented  by Band function \citep{Band_1993}. In this section
we will also investigate how the scattered photon contribution may change the
spectral characteristics and possibly contribute artefacts in the exact
evaluation of the spectral slope, if they are not accounted for correctly. We
conclude in Section~\ref{sec:discussions} by discussing about the outcome of our study and describing our future plan of work.

\section{Monte Carlo simulation with GEANT4}\label{sec:simulation}
We have used GEometry ANd Tracking 4 (GEANT4) for simulating the Compton reflection of X-rays from Earth's atmosphere.
GEANT4 is a well known Monte Carlo detector simulation package \citep{Agostinelli2003} written in C++, initially developed
for the simulation of high energy physics and gradually got enhanced, in order to be applied to lower energies also.
Currently, it is used extensively in high energy detector and instrument simulation (mass model) of most of the
astronomical missions including \asat-CZTI \citep{Mate2021}.
%We are currently in the process of incorporating the atmospheric scattering response to  data processing, in which the instrument mass-model  is itself constructed in GEANT4, so seamless amalgamation of the two is expected.
Below we describe the construction of the
simulation geometry including the Earth's atmosphere, the generic detector at LEO height, the physics models used
in the simulation, and the methods for particle generation, data extraction, and analysis.

\subsection{Geometry and Earth's atmosphere}\label{sec:geometry}
The Monte Carlo code developed for this study primarily imitates the interactions of incident parallel mono-energetic
beams of X-ray and \gray\ photons with Earth's atmosphere and find the detection responses at an imaginary flat
and fully efficient detector placed at LEO, at height of 650~km.

The attenuation of incoming X-ray to \gray\ photons in the atmosphere occurs predominantly from few 100s of km down to $\sim$10 km \citep[see][]{Palit2018_1}. About 99.7\% of Compton back-scattering events
occur below 125~km \citep{Willis2005},
with the majority of the interactions occurring in the 20--50~km range.
Considering this, we restrict the atmospheric heights from $\sim$10~km to 500~km in the simulation. The lower part of this range is of
significantly varying density and concentration of neutral components. Those are mainly Nitrogen molecules (N$_2$),
Oxygen molecules (O$_2$)  and some \rough{other less abundant} elements like Helium and Argon, which play an important role in the interaction process
of high energy photons. Above $\sim100$~km the mass column density of the atmosphere is below $1.4 \times 10^{-3}~\mathrm{g~cm}^{2}$
and the optical depth is so low that the variation in such concentration has negligible effect. Other than the inhomogeneity
along vertical direction, the atmosphere also has inhomogeneities along latitude and longitude, and is dynamic in nature.
For example, the atmosphere is thinner over the poles and thicker over the equator.
Since our focus is on satellites in Equatorial orbits and we aim to calculate the nominal scattering effect, these are
%We are interested in estimating the atmospheric Compton reflection mostly below 90 degree zenith angle and assume equatorial LEO orbit, so, these
secondary concerns and can be neglected safely. In view of this we incorporated only vertical stratification in the
atmosphere with the necessary data on density, and various neutral concentrations.

\geant\ detector or geometry construction supports only concentric spherical shells. We model the Earth's atmosphere using multiple layers in the detector construction class. The distribution of layers is as follows: The first ninety layers starting from 10~km 
above the Earth's surface have a thickness of 1~km each, followed by 20 layers of 5~km each, and lastly the 12 outermost layers are of 25~km each. They are
formed with average molecular densities and other atmospheric parameters at those heights. Corresponding neutral atmospheric
data are obtained from NASA-MSISE-90 atmospheric model \citep{Hedin1991}. This atmospheric structure is directly imported
from previous studies \citep{Palit2013, Palit2018_1}, which already have successfully evaluated the X-ray interaction
process though in the context of solar flares and their effect on radio propagation. A NaI detector disc of radius of
100~km and thickness of 10 meter (ensuring 100\% absorption of received photons) is placed at a height of 650~km from the Earth's surface
at the zenith (direction, vertically upward from Earth's centre).  A schematic of the simulation geometry
along with the photon generation and propagation through atmosphere has been presented in Figure~\ref{fig:geom}.
%\todo{check: are NaI detector effects (for instance energy resolution, escape peaks etc) modeled?}

\subsection{Physics of interaction}
The main interaction process for energetic photons corresponding to soft to hard X-rays and soft \gray s with
the atmospheric molecules are the primary photo-ionization and subsequent secondary electron ionization. The ionization
can be due to photoelectric effect and Compton scattering. Electron-positron pair creation takes place for photon energies
exceeding $\sim$ 1000~keV and has very small contribution in the overall interaction statistics. Due to the presence of
a small amount of Helium and few other heavier elements in the Earth's atmosphere The photoelectric cross section exceeds the
scattering one up to energies $\sim$30 keV and inelastic (Rayleigh) scattering in atmospheric composition dominates the
total scattering cross section at energies below 10--20~keV \citep{churazov2008}.

Each of the Physics processes in GEANT4 uses several models. Models for all electromagnetic processes can be subdivided into four 
general physics scenario categories: standard, low-energy, polarized and adjoint. For unpolarized low energy photons, most Compton 
scattering models corresponding to `low-energy' and the `standard' categories use the Klein-Nishina approach \cite{1929ZPhy...52..853K} 
for cross-section. The other two efficient models (of 'low-energy' category), namely, Livermore and Penelope based Compton scattering 
models  use cross section calculation based on the EPDL data library \citep{CULLEN1995499} and Penelope code \citep{SEMPAU2003107}. 
All these models, namely, the \sw{G4LivermoreComptonModel}, \sw{G4PenelopeCompton\-Model}, \sw{G4LowEPComptonModel} etc. produce 
very similar outcome for our simulations. In the subsequent sections, we present the results computed with the \sw{G4PenelopeComptonModel} 
and corresponding other Penelope low energy physics models.

\subsection{Source photons, interactions and simulation steps} \label{sec:simdetails}

\begin{figure}
\centering{
\includegraphics[height=3.0in,width=3.0in,angle = 0]{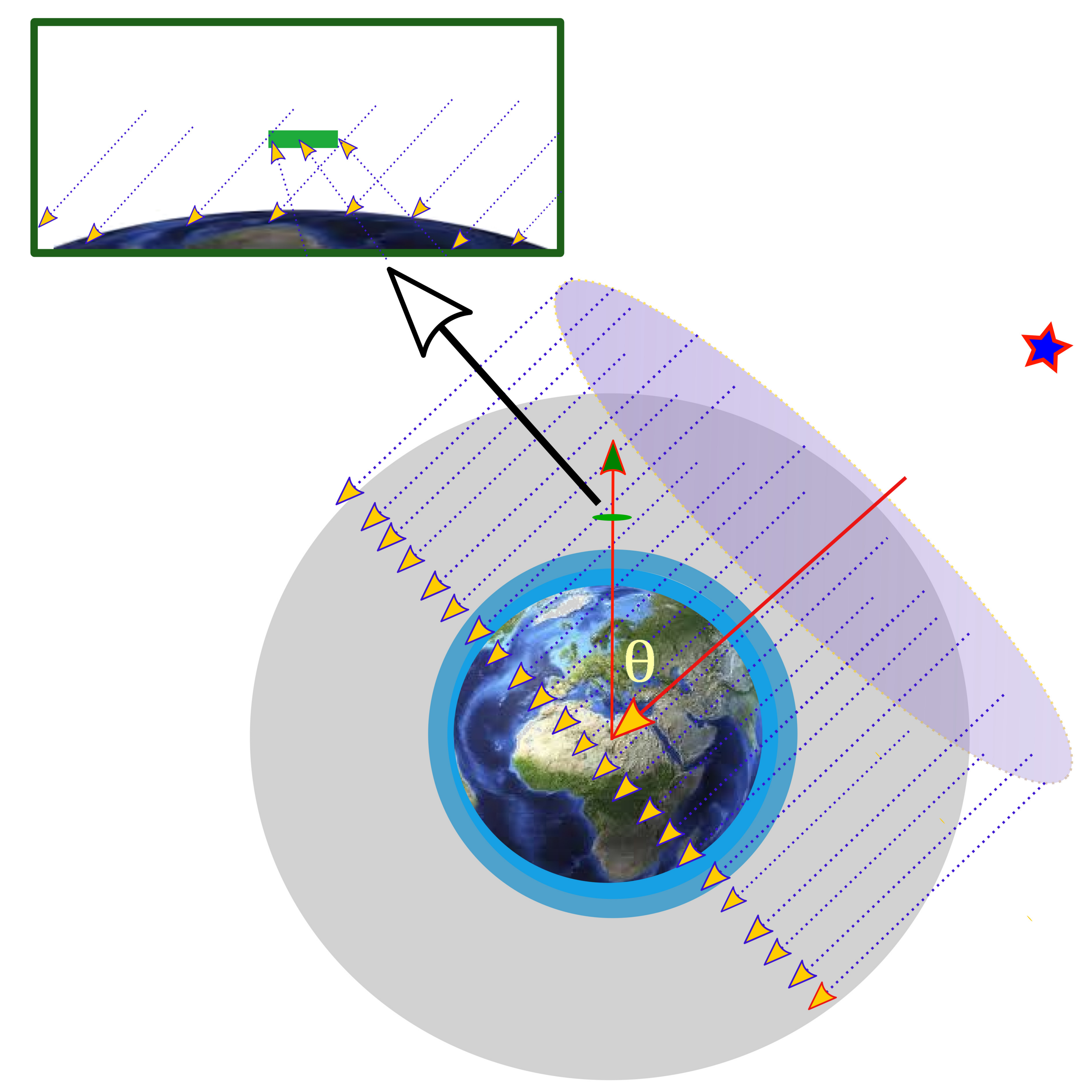}
\caption{A schematic of Geometry set up for Monte Carlo simulation consisting of Earth's atmosphere, the ideal detector (green) at
LEO height, parallel beam of photons originating from a flat circular disc of radius 7200~km touching the surface of
an  Earth-centric imaginary sphere of same radius and falling on the atmosphere of Earth is shown.}
\label{fig:geom}}
\end{figure}

During source photon generation within primary generation class of \geant, an imaginary hollow sphere of large radius (7200~km)
is assumed  concentric to the Earth's layered atmosphere. The photons are generated randomly on the surface of an imaginary
flat disc (Figure~\ref{fig:geom}) of same radius, touching the outer surface of the sphere at a point of intersection of the
incoming photon direction with it and placed tangentially. This ensures that the photon incidence corresponds to plane wave
coming from virtually infinite distance. During the simulation the detector is always kept at the vertically up-word
position (zenith) that is on an imaginary vertical line extending upward from the center of the Earth. The angular position of
the source ($\theta$) i.e., the incidence angle is always set with respect to the vertical line (Figure~\ref{fig:geom}). $\theta = 0\degr$ corresponds to normal incidence on the atmosphere, with the detector directly between the source and the Earth. The detector is insensitive to direct source photons. The effect of the small shadow of the detector is corrected for by scaling up the observed counts to the unobstructed cross-section of the Earth.

We run each of our simulations
with a mono-energetic photon beam having sufficient number ($10^8$) of photons, so that the detector can collect $\sim$10000
scattered photons from the whole of the atmosphere. Each run comprises of primary photons at a single energy. Our simulations span the energy range from 5~keV to 5~MeV as follows: 2~keV intervals from 10 to 50~keV, 5~keV intervals from 50 to 200~keV, 20~keV intervals
from 200 to 400~keV, 100~keV intervals from 400 to 1000~keV, and simulations at 2000, 3000, 4000 and 5000~keV. This division is carefully
chosen to get a smooth Atmospheric Response Matrix (ARM) (described in \S\ref{sec:results}) by interpolation of the resultant
mono-energetic distribution functions over incident energy values.
% to get the same for each and every keV values in the whole input energy interval.
Simulations are carried out for varying incidence angles ($\theta$), and here we focus on a representative subset of these incidence angles. Details of the simulation steps and
subsequent processing and analysis are covered in
the \S\ref{sec:results}.

\section{Results}\label{sec:results}
As discussed in the previous section, we undertake simulations only for mono-energetic beams with different incidence angles. The outcome, with suitable interpolation can easily
be converted into response matrix and the spectrum for any realistic 
astrophysical source has to be synthesized by using a weighted sum of the simulated mono-energetic responses. The weights are to be determined by the number of source photons
expected at that incident energy. The exercise of starting with mono-energetic
incident beams has two main advantages. Firstly, it is helpful in better
understanding of
the exact nature of the atmospheric scattering response and how it varies over the incident photons at different ranges of X-ray/\gray\ energies. Secondly, this allows us to re-use our simulation results for any incident spectrum, without resorting to computationally expensive runs for each of them.

\subsection{Mono-energetic response and Atmospheric Response Matrix (ARM)}
\begin{figure}
\centering{
\includegraphics[width=\columnwidth]{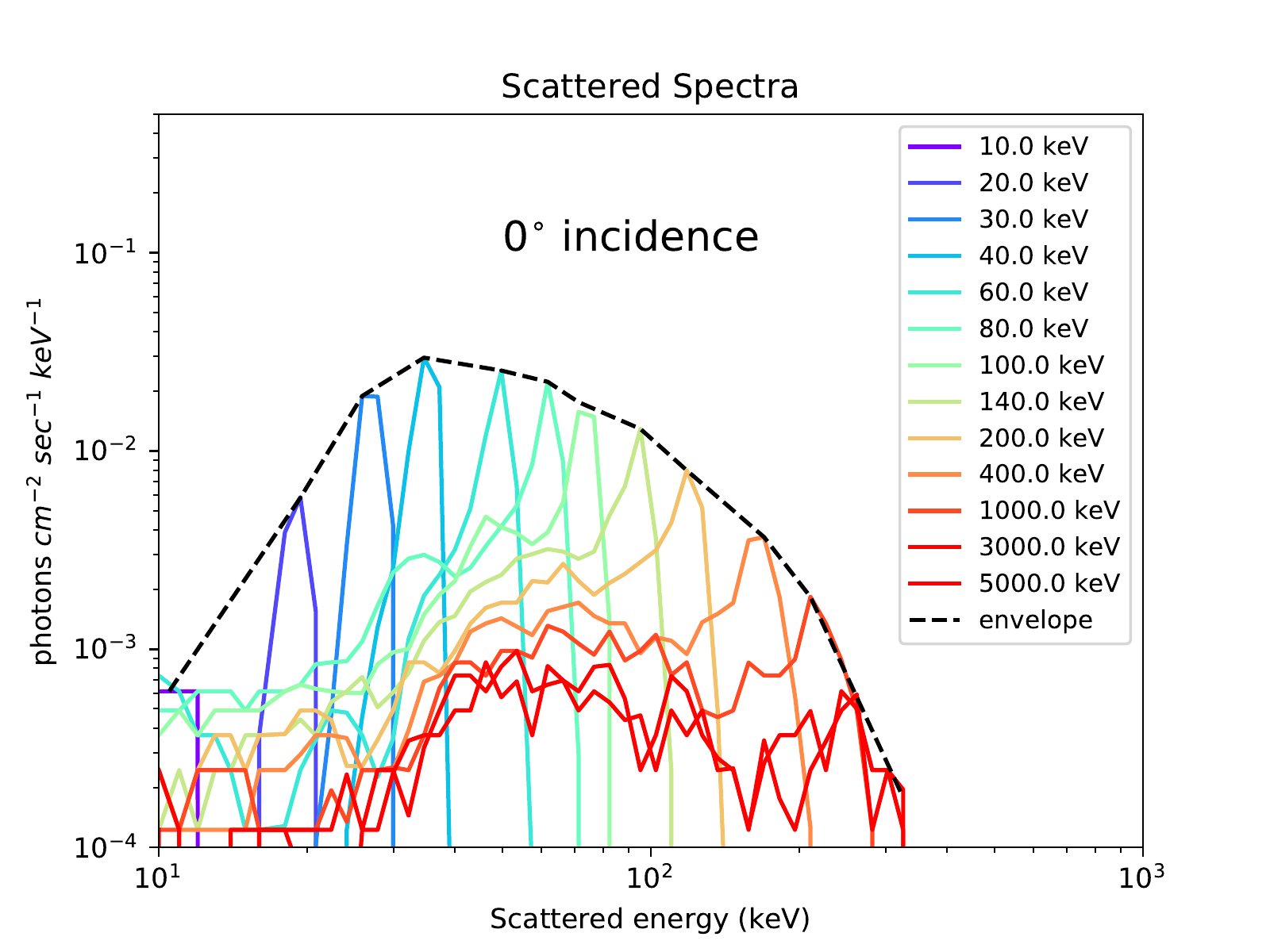}
\caption{Calculated response functions (spectra) are shown in the plot for 0\degr\ incidence angles of monochromatic photon beams. Here, the horizontal axis corresponds to the detected atmospheric scattered photon energy. The vertical axis represents the number of reflected (scattered) photons from Earth's atmosphere detected at 1 $cm^2$ of detector area at the detector at LEO height due to 1 incidence photon per square of centimeter of whole of the Earth's atmosphere
at the mentioned energies and angle. The dashed black line denotes the overall envelope of the responses.}
\label{fig:mono1d}}
\end{figure}

Figure~\ref{fig:mono1d} shows the detected (simulated) atmospheric scattered spectra for few of the incident energies,
normalized for a single incident photon at that energy. The incidence angle ($\theta$) for this plot is 0\degr. Due to inelastic interactions, many source photons get scattered to lower energies, creating a broadband spectrum for each of the mono-energetic incident photons. The maximum energy of scattered photons, as well as the energy at which the scattered spectrum peaks, are both below the incident energy.
Both the parameters increase with the incidence beam energy up to few hundreds
of keVs, but as incidence energy goes higher and crosses 1 MeV the maximum and
peak energies of scattered photons stop increasing. We found that whatever
higher the incident energy values are used, the energies of scattered photons remain
well below $\sim$1 MeV.  The maximum value of the scattering response peak
occurs for incidence photon energy of $\sim$40 keV and lies at
detected energy of $\sim$35 keV.
Note that this maximum refers only to the peak in the observed spectrum, and
not the total number of detected photons. The response diminishes as one goes
from lower incidence angles to higher ones. 

\begin{figure}
\centering{
\includegraphics[scale=.30]{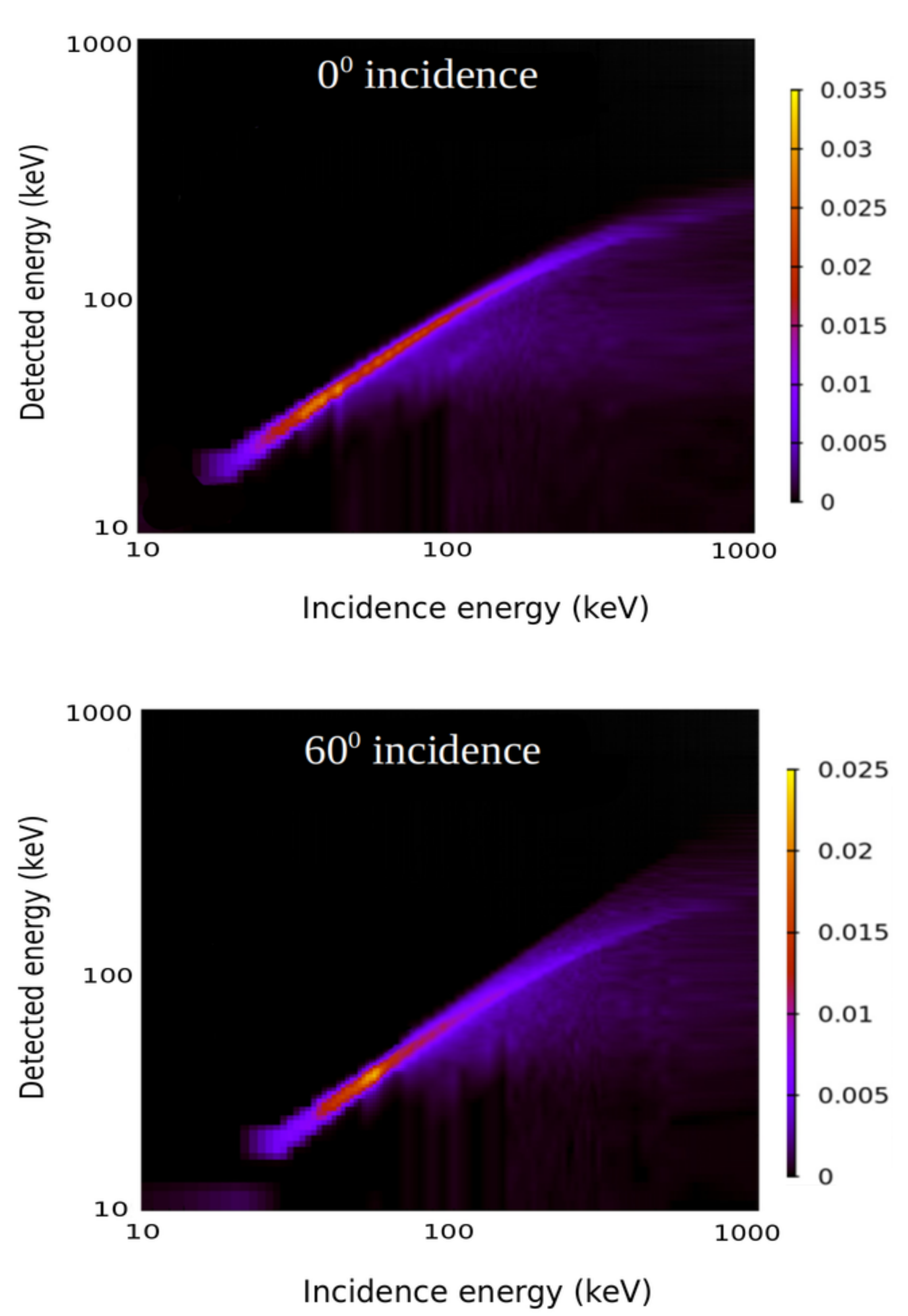}
\caption{ARM for two angles of incidence. Simulations are carried out at various incidence energies as discussed in Section~\ref{sec:simdetails}, and are interpolated for other energies. The top panel is a two-dimensional version of the plots in Figure~\ref{fig:mono1d}, for photons incident at $\theta=0\degr$, while the bottom panel is the ARM for $\theta = 60\degr$.}
\label{fig:mono2d}}
\end{figure}

In Figure~\ref{fig:mono2d}, the detected atmospheric responses to incidence
photons of different energies are demonstrated for two different angles of
incidence with values represented by the color bars. The $Y$-axis corresponds
to the energy of the incident photons, while the $X$-axis denotes the energy of
scattered photons. The color coding shows the intensity at each energy pair. It
is obvious that the detected energy is never greater than the incident energy,
leaving the \rough{upper left} part of these figures blank. These plots represent the
ARM for various incidence angles, and the curves shown in
Figure~\ref{fig:mono1d} correspond to horizontal slices through the 
plot in top panel of Figure~\ref{fig:mono2d}.
Values for incident energies other than those used in our simulation grid are
obtained by suitable interpolation.

\begin{figure}
\centering{
\includegraphics[scale=0.35]{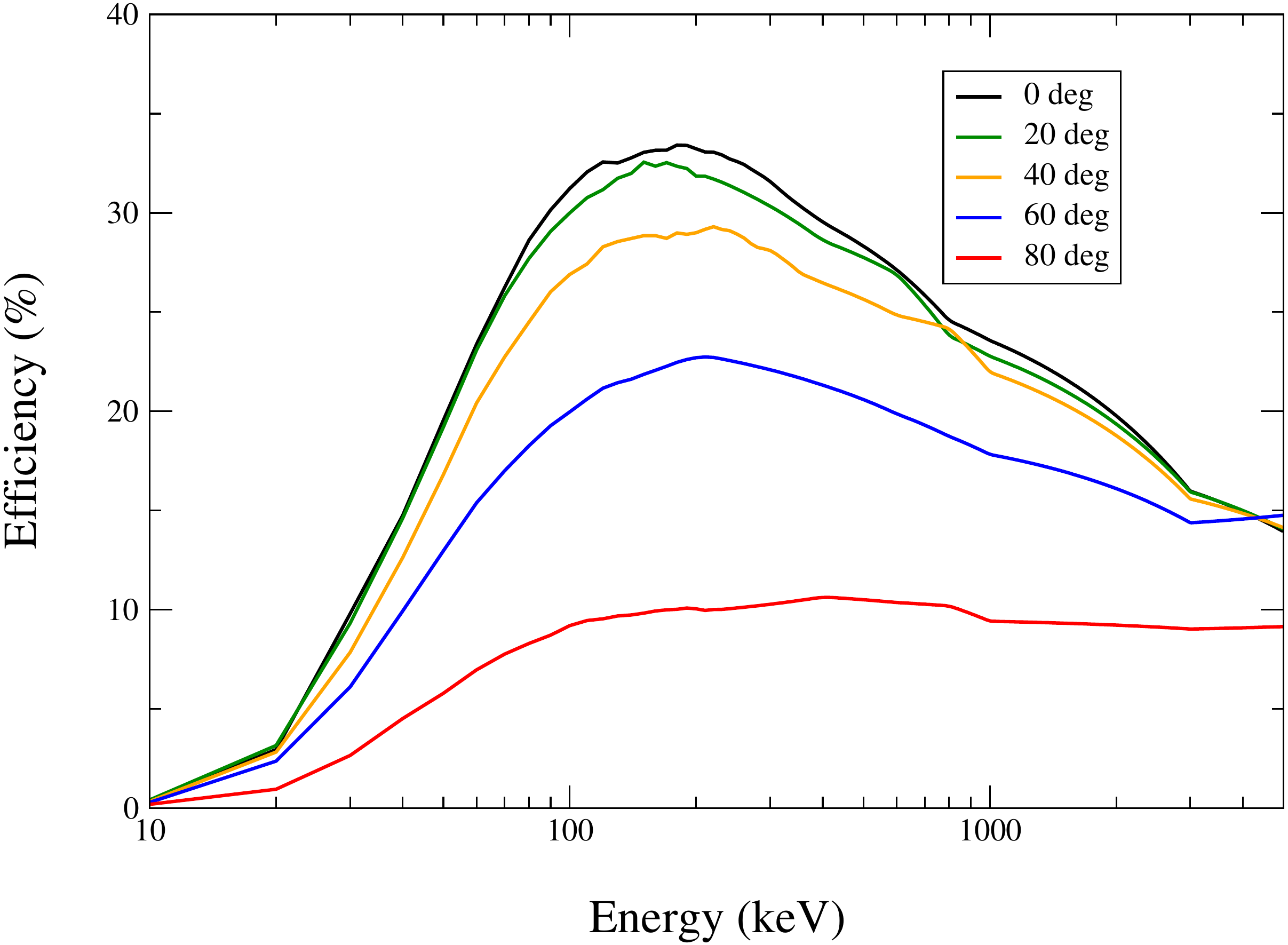}
\caption{Efficiency (ordinate) of Compton reflection of the incident photons as function of incidence energy (abscissa)
for various incidence angles.}
\label{fig:eff}}
\end{figure}

For each energy of incidence, we calculate the area under the scattering response curve (Figure~\ref{fig:mono1d}) to obtain the atmospheric scattering efficiency, i.e., the total number of scattered photons (at any possible scattered energy) for a 
single incident photon
on Earth's atmosphere. We converted this in percentage and plot in Figure~\ref{fig:eff} as function of incidence energy
for various incidence angles. The plots show that the scattering efficiency is maximum for 0\degr\ incidence of photons (black curve)
and diminishes as the angle of incidence increases. For normal incidence the highest efficiency is at $\sim120$ keV, and as Figure~\ref{fig:mono1d} shows, these photons are spread over a broad band of lower energies.
The energy at which the atmospheric scattering efficiency is highest tends to gradually increase until the efficiency profile becomes nearly uniform in all energies above $\sim$100 keV for large incidence angles.

\subsection{Spectral response for GRBs}\label{sec:spec_grb}

We now discuss the scattering response of the Earth's atmosphere to an incident GRB spectrum, approximated by a typical Band function \citep{Band_1993}. The parameters of the Band function given by \cite{2015MNRAS.448.3026W} corresponding to a typical short GRB, viz, $\alpha = -0.5$, $\beta = -2.25$, and $E_\mathrm{peak} = 800$~keV are chosen for the calculation. The scattered component corresponding to the spectrum for 0\degr incidence is estimated by the convolution of the spectrum with the corresponding response matrix (\texttt{ARM} representing the atmospheric scattering  in top of the Figure~\ref{fig:mono2d}). We pick the norm for the band function such that the fluence of the GRB is $10^{-6}\mathrm{~erg~cm^{-2}}$(20-200 keV) over 1 sec interval of prominence.

In order to disentangle the effects of the ARM from any instrument response effects, we assume that the instrument is an ideal omnidirectional detector: it measures the exact energy of each incident photon, but cannot discern photon incidence directions. In actual analysis of course, the direction- and energy-dependent response of the instrument would have to be folded into the calculation.

%We consider the total detected spectrum as the sum of a chosen incident spectrum and the corresponding scattering spectrum. We then fit this total spectrum with a model consisting of incident and a thermal spectrum to demonstrate the apparent resemblance of the scattered component as a thermal spectrum. We then go on to find out the relative strengths on the individual components to find out how strong the Earth scattering can be. While we realise that no one actually forgets to include the Atmospheric Scattering Response in the model fitting, we then go on to show, in \S\ref{sec:inexact_ARM}, how systematic errors in the Atmospheric Response Matrix (ARM) can affect the modelled spectrum.
%For long GRB, the spectrum defined by a Band function with parameters, $\alpha = -1$, $\beta = -2.25$, and $E_\mathrm{peak} = 511$ keV \citep{2010MNRAS.406.1944W} is considered.
%Exactly same spectral parameters are adopted for the spectral comparison of CZTI mass model (\todo{Mate et al., this issue}).
 
%We refer to the two short GRB simulations for $\theta=0\degr$ and $\theta=60\degr$ as GRBS0 and GRBS60 respectively, with the corresponding long GRB simulations being referred to as GRBL0 and GRBL60.

\begin{figure}
\centering{
\includegraphics[scale=.5]{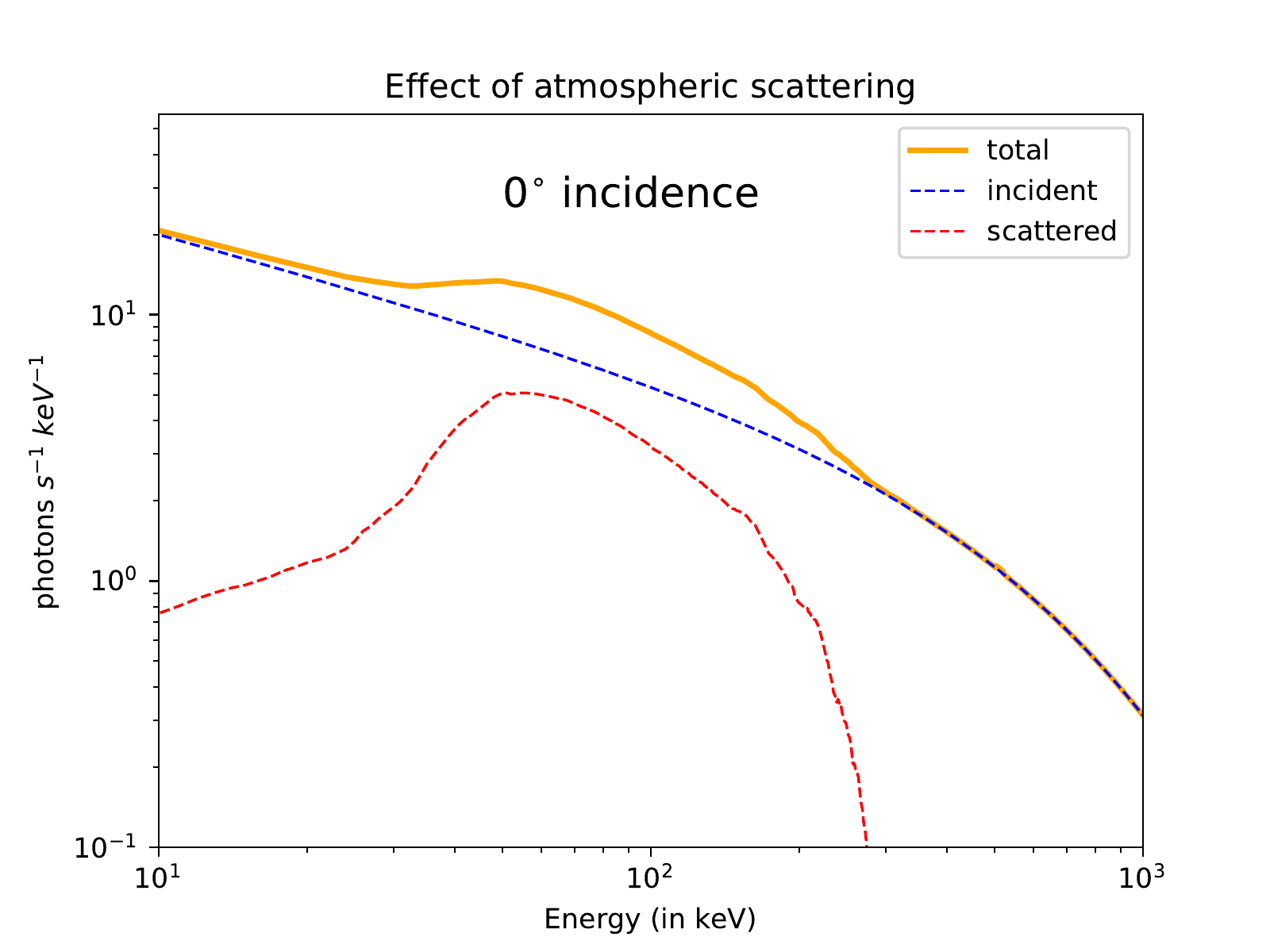}
\includegraphics[scale=.5]{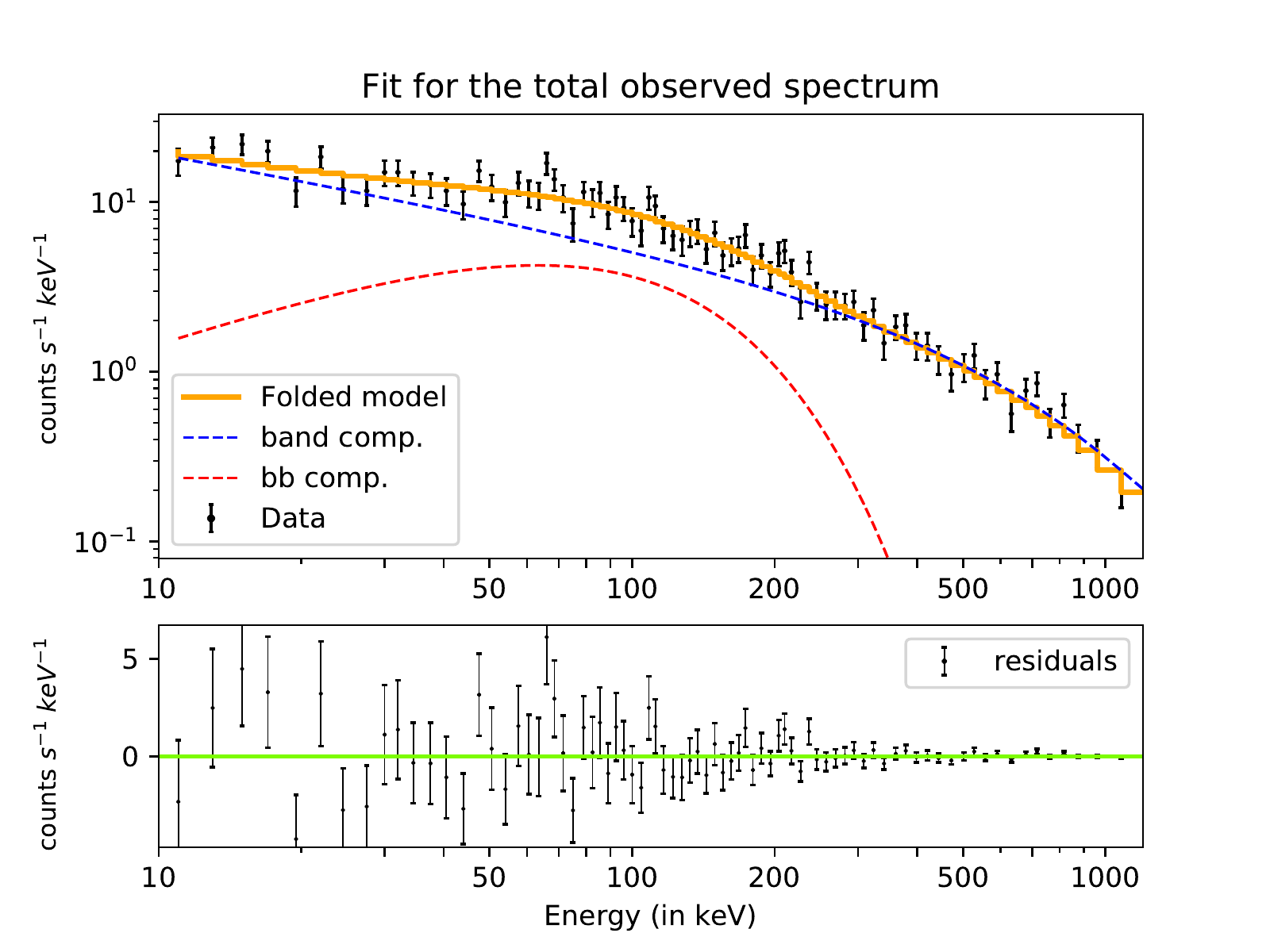}
\caption{Simulations of reflected spectra (Top panel) and \xs\ fits to the data (Bottom panel) for the GRB considered are demonstrated. The dashed blue line in the top panel shows the spectrum of a short GRB with parameters $\alpha = -0.5$, $\beta = -2.25$, and $E_\mathrm{peak} = 800$~keV, which is incident at 0\degr and detected directly. The dashed red line shows the scattered spectrum. The solid orange line denotes the net spectrum incident on the detector. The bottom panel shows the simulated photon data and the best-fit model obtained from \xs. Black symbols with error bars show the simulated data from the grouped \sw{.pha} files. Dashed blue and red lines show the best-fit Band (\sw{grbm}) and blackbody (\sw{bbody}) components used in the fitting, respectively. The solid orange line is the sum of these two components, which agrees well with data giving residuals consistent with zero. The best-fit parameter values are given in Table~\ref{tab:fit_param}.}
\label{fig:short_grb}}
\end{figure}

Simulation results for the atmospheric scattering component is shown in the upper panel of Figure~\ref{fig:short_grb}. The dashed blue curve is the incident band spectrum, that would be detected directly, and the dashed red curve shows the scattered component from the atmosphere. The solid orange curve represents the net detected spectrum by a detector in LEO. We can see that a significant amount of photons scattered by the atmosphere are detected at the intermediate energy range of $\sim$ 30--300~keV. This reflected component produces a broad hump around $\sim$60 keV on top of the direct Band spectrum.

\begin{table}[ht]
\centering
\small
\begin{tabular}{|*{2}{c|}}
\hline
{\textbf{Energy range (keV)}}  & {\textbf{Flux ratio}}\\
\hline
50 -- 100 & 0.66 \\[3pt]
\hline
100 -- 200 & 0.46 \\[3pt]
\hline
200 -- 400 & 0.08 \\[3pt]
\hline
 400 -- 1000 & 0.001 \\[3pt]
\hline
20 -- 2000 & 0.25 \\
\hline
\end{tabular}
\caption{Ratio of detected atmospheric scattered flux to that of incidence flux integrated over various energy ranges}
\label{tab:tab1}
\end{table}

In Table~\ref{tab:tab1} we show the ratio of the detected scattered flux from atmosphere to that of the incident flux integrated over various energy ranges for the Band function corresponding to the GRB used in our study. We see that the number of reflected photons in the 50--100~keV range is as high as 66\% of the incident photons, which corresponds to $\sim40\%$ of the total observed photon counts in this energy range. 

At a glance, the scattered component mimics a blackbody component (Figure~\ref{fig:short_grb}, Top panel), and would have to be modelled so, if atmospheric scattering effects were not accounted for. As an illustration, we show such a naive analysis where the net observed spectrum can indeed be fit well by a Band + blackbody spectral model (Figure~\ref{fig:short_grb}: Bottom panel).

To undertake formal fits we simulate an ``observed'' spectrum 
consisting of the direct incident and reflected spectra. The \geant\ output is
binned at 1~keV resolution, and multiplied by the effective area of the
detector to obtain the expected count rate $\lambda_E$ (in units of
counts~keV$^{-1}$) as a function of energy.
%\rough{As mentioned above, we assume an ideal detector where the effective area is constant independent of energy, and there is energy redistribution (the response matrix is diagonal).}
To account for Poisson noise, we
calculate the final observed number of photons in each 1~keV by using a Poisson
distribution with rate equal to $\lambda_E$. The uncertainty for each bin is
set to $\sqrt{\lambda_E}$. We note that in practice, the uncertainties will be
higher due to the presence of background counts. Both the spectrum and
uncertainty are written into a \sw{.pha} file with 1~keV channels, spanning the
energy range from 10~keV -- 2~MeV.  We assume an ideal detector of 150~cm$^2$
collecting area, comparable to several operational GRB instruments. We assume
that the effective area of the detector is independent of energy, and that the
detector reports the true energy of each incident photon without any
redistribution. The detector response (\sw{.rsp} file) is then modeled simply
as a diagonal matrix, with diagonal elements equal to the effective area.

\begin{table}[!thbp]
    \def\arraystretch{1.5}
    \centering
    \begin{tabular}{|c|c|c|}
    \hline
    Model name & Parameter & Value \\
    \hline
    \multirow{4}{*}{Band} & $\alpha$ &  -0.51 $\pm$ 0.06  \\
    & $\beta$ & -2.4 $\pm$ 0.5  \\
    & $E_{c}(keV)$ & 560 $\pm$ 83  \\
    & $norm/10^{-3}$ & 40.3 $\pm$ 3.6  \\
     \hline
    \multirow{2}{*}{Black Body} & $kT(keV)$ & 39.7 $\pm$ 3.4  \\
    & $norm$ & 8.6 $\pm$ 1.6 \\
     \hline
    \multicolumn{2}{|c|}{\rough{$\chi^2$ (DOF)}} & \rough{81 (76)} \\
     \hline
    \end{tabular}
    \caption{Best-fit values for all the parameters with their 1-$\sigma$ error and the reduced $\chi^2$ of the fit presented in bottom panel of Figure~\ref{fig:short_grb}. Description of the parameters: \texttt{$\alpha$} and \texttt{$\beta$} are Band spectral indices, $E_c$ is the characteristic energy; related to peak energy $E_p$ by $E_p$ = $E_c$ (2 + $\alpha$). \textit{T} represents the temperature of the black body where \textit{k} is the Boltzmann constant. Both $E_c$ and \textit{kT} are expressed in keV. \textit{norm} is the spectral norm for each of the components of the model and is expressed in $\mathrm{photons~cm^{-2}~s^{-1}~keV^{-1}}$.}
    \label{tab:fit_param}
\end{table}

We then follow usual X-ray data analysis procedures. We run the \sw{grppha}
\sw{FTOOL} on the \sw{.pha} file to create channel bins having at least 30
photons each. We then attempt to fit the \sw{grbm} spectral model to this
simulated data in \xs. As expected, the fits are of poor quality due to the
presence of the reflection hump in the data. We then model the data with
\sw{grbm + bbody}, and obtain acceptable fits (Table~\ref{tab:fit_param},
Figure~\ref{fig:short_grb}: Bottom panel).

In this naive analysis, where the fitting process ignores the atmospheric scattering effects, we find that the best-fit blackbody temperature is $kT\sim40$~keV. 
The spectral parameters $\alpha$ and $\beta$ are satisfactorily recovered. 
The \xs\ \sw{grbm} model parameterizes the Band function in terms of the 
characteristic energy $E_c$, which is related to the peak energy \
as $E_p = E_c (2 + \alpha)$. Our simulation thus uses $E_c =  533$~keV. 
We find that the best-fit value is consistent with the simulation, albeit 
with large error bars. We attribute these uncertainties to small number 
statistics: with our selected parameter values, we have $\sim 1000$ photons, 
with only a small fraction of those incident at energies above $E_c$.

To put these results in perspective by comparing to a real detector, we proceed
to measure the possible contribution of both components to the flux in the
\asat-CZTI 20-200~keV band. 
%We note that the 20--200~keV incident fluence and hence the flux for the 1 second-long short GRB considered is $10^{-6}~\mathrm{erg~cm^{-2}}$.}
The flux of individual components is calculated
by using the \sw{cflux} model in \xs. We find that the best-fit black body
contributes about a third of the total flux in the energy range of interest
(Table~\ref{tab:flux}).

\begin{table}[!th]
    \centering
    \def\arraystretch{1.5}
%    \begin{tabular}{|c|c|c|}
%    \hline
%%    \hline
%    \multicolumn{3}{|c|}{20--200~keV Flux (in units of $10^{-7}$ $\mathrm{erg~cm^{-2}~s^{-1}}$)} \\
%    \hline
%     Total & Band (\% total) & Black body (\% total) \\
%     \hline
%     $9.9^{+1.0}_{-1.0}$\ \  (63.8 \%)& $5.4^{+1.4}_{-1.3}$\ \  (36.2 \%) & $15.3^{+1.78}_{-1.6}$ \\
%    \hline
%    \end{tabular}
    \begin{tabular}{|p{2cm}|c|c|c|}
    \hline
%    \hline
    \multicolumn{4}{|c|}{20--200~keV Flux} \\
    \hline
    & Band & Black body & Total \\
     \hline
     Flux ($10^{-7}$ $\mathrm{erg~cm^{-2}~s^{-1}}$) & 
     $9.9^{+1.0}_{-1.0}$ & $5.4^{+1.4}_{-1.3}$ & $15.3^{+1.78}_{-1.6}$ \\
     \hline
     Fraction of Total & 65\% & 35\% & 100\%  \\
     \hline
    \end{tabular}
    \caption{Comparing the contributions of incident and scattered components in the total GRB flux measured by a detector like \asat-CZTI. Flux values are reported in the 20 -- 200~keV range, in units of $10^{-7}$ $\mathrm{erg~cm^{-2}~s^{-1}}$. Numbers in parenthesis indicate the percentage contribution to the total flux. Fluxes of individual components are measured by using the \sw{cflux} model component in \xs. Error bars denote and 90 \% confidence intervals. The relative contributions of the individual components to the total flux are mentioned as (\% total).}
    \label{tab:flux}
\end{table}

In practice, of course, analyses of GRB spectra account for the atmospheric contribution -- often transparently to the end users. For example BATSE and \fermi\ have introduced the correction in their analysis step by convolving ARM to the Detector Response Matrix (DRM) \citep{Pendleton_1999, Meegan_2009}. The ARM is obtained primarily by performing extensive Monte Carlo simulation of interaction of incident photons with the atmosphere and then improving on it gradually with some observation of the same for X-rays from solar flares in an iterative manner.

However, It is obvious from above exercise, that the unchecked atmospheric contribution can be misinterpreted as an added thermal component during fitting of observed counts, though the odd of it to actually happen is small, as most, if not all the missions already or will have the provision of  mitigating the effect through the inclusion of ARM. the similarity of this effect to blackbody spectra raises the question of the susceptibility of models to errors in the ARM. We explore this further in \S\ref{sec:inexact_ARM} by testing the effects of $\sim 10$\% errors in the \texttt{ARM} on the spectral analysis.

We note that throughout this work, we consider the ideal detector case. In practice, the effect of such errors on the final fit will depend on the direction-dependent sensitivity of the instrument. In a case where the instrument has higher on-axis sensitivity as compared to off-axis sensitivity, the reflected component will be weaker if the instrument was pointing to the GRB, and stronger if the instrument was pointing towards the Earth. An illustration of the latter case is when the primary target of CZTI is almost occulted by the Earth, when the boresight points straight to the atmosphere. In such ``strong reflection'' cases, even smaller errors in the \texttt{ARM} can cause large changes in the inferred parameters.

\subsection{Effects of imcorrect ARM}\label{sec:inexact_ARM}

In order to demonstrate how systemic errors in the \texttt{ARM} can affect the modelled spectrum, we synthesise three scenarios. In the first one, we show that we can recover the true incident spectrum by accurately accounting for the \texttt{ARM} (100\%). In the other two, we introduce a $\mp$10\% error to the \texttt{ARM} and then use these incorrect response matrices as the atmospheric responses to demonstrate the effect of improper modelling of atmospheric response. We examine the three scenarios for two types of sources (incident at 0\degr angle): a GRB without a blackbody component (\S\ref{sec:pureband}), and a GRB with an intrinsic blackbody component (\S\ref{sec:grbmbb}). In order to get enough photon statistics, we consider a GRB with 50~s duration. The spectral parameters for the Band component are same as in \S\ref{sec:spec_grb}.

We simulate observations using the same methods as discussed in \S\ref{sec:spec_grb}, hence the observed spectrum consists of the incident GRB spectrum as well as the photons scattered from the atmosphere. While analysing the data for both source models, we include the \texttt{ARM} as per the three cases: the exact ARM, an underestimated \texttt{ARM} (corresponding to 90\% reflection strength), and an overestimated \texttt{ARM} (corresponding to 110\% reflection strength): for simulating the $\mp$10\% systematic errors.

\subsubsection{Band spectral model:}\label{sec:pureband}

\begin{figure*}[htbp]
    \hspace{-2cm}\includegraphics[scale=0.7]{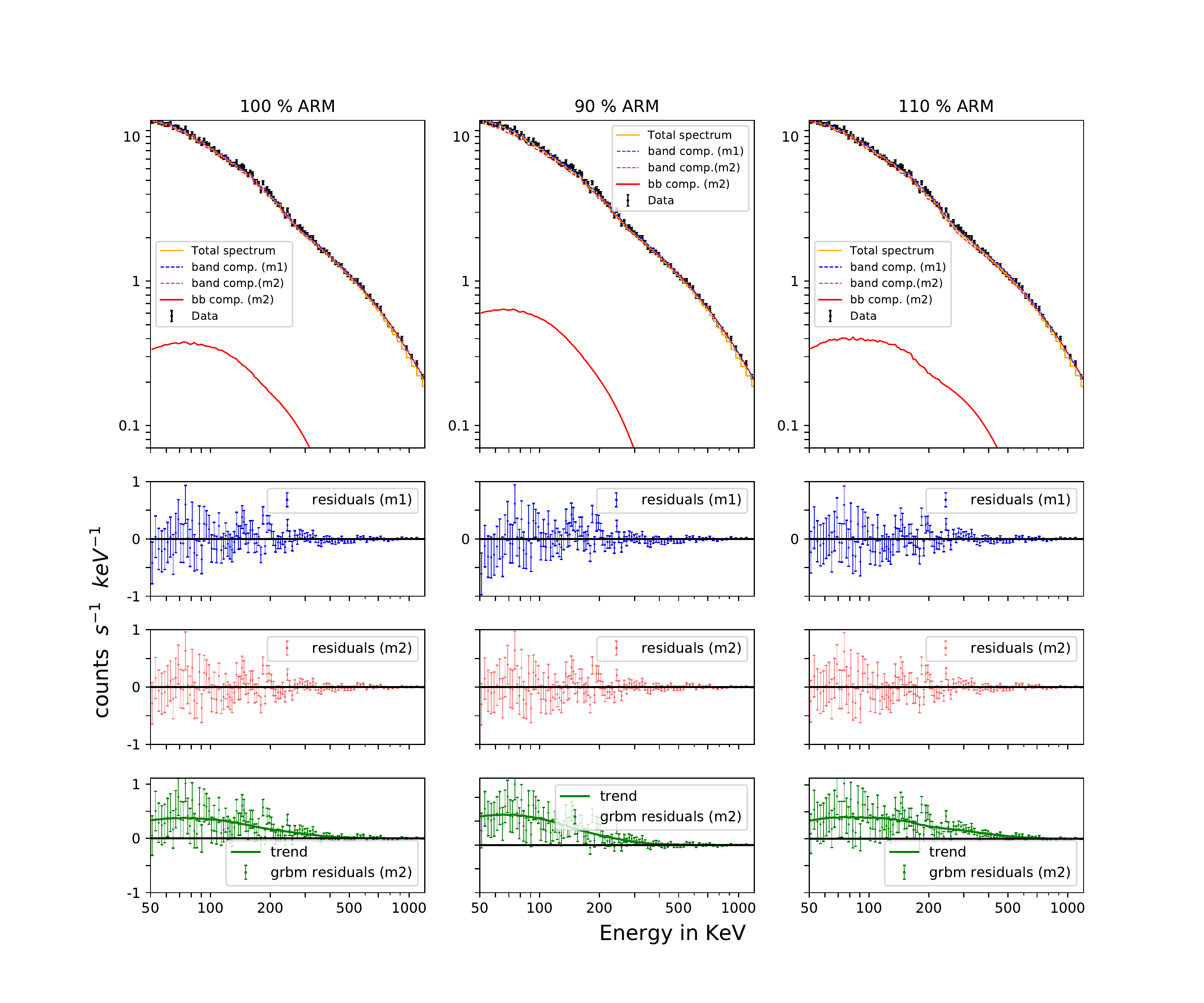}
    \caption{Simulations of the total GRB spectrum (incident+scattered) and the corresponding \xs\ fits to the simulated data for Band incidence. Each column corresponds to a different response matrix (Atmospheric Response Matrix (\texttt{ARM}) convolved with the detector response matrix (\texttt{DRM})). The first column represents the accurate \texttt{ARM} and the next two columns correspond to \texttt{ARM} which is under and over estimated by 10\%. The incident spectrum is that of a GRB modelled by a band function with parameters $\alpha = -0.5$, $\beta = -2.25$, and $E_\mathrm{peak} = 800$~keV ($E_c=533$~keV) and \textit{norm} = 0.0427. The simulated data are marked in black (with error bars overlaid). The solid orange line denotes the \xs\ fit for the simulated data. Two models are used to fit the data. Model 1 (m1) is the band function (\texttt{grbm}) and model 2 (m2) is band function (\texttt{grbm}) with an additive blackbody (\texttt{bb}) component. The top row shows the \xs\  fits to the simulated data with m1 represented in blue and m2 depicted by red colour \rough{(changes in the band component are difficult to observe directly when plotted in the absolute scale and hence successive residual plots are presented to depict the effect of the two models considered)}. The second row from top shows the residual plot (blue) when the data is fit with m1. The third row from top shows the residuals (red) when m2 is used to fit the data. The bottom row shows the residuals (green) of the \texttt{grbm} component of model m2. These are obtained by fitting the whole model m2 to the data and then deleting the \texttt{bb} component, thus highlighting the presence of a residual black body (if any), \rough{with the black body component itself overlaid marked as \textit{`trend'}}. The best-fit parameter values are given in Table~\ref{tab:band_sc_fit_param}.}
    \label{fig:band_sc}
\end{figure*}

In the ``Pure Band'' incidence case, the source spectrum is represented by a Band function, with 
the parameters $\alpha=-0.5$, $\beta=-2.25$, $E_{peak}=800$~keV ($E_c=533$~keV) and \textit{norm}=0.0427.
Analysis results are shown in Figure \ref{fig:band_sc}, with best-fit parameters given in Table~\ref{tab:band_sc_fit_param}. The topmost panels show the simulated data in gray, along with the folded model in orange. The three columns show the cases where the analysis uses the correct \texttt{ARM}, an underestimated \texttt{ARM} and an overestimated \texttt{ARM} from left to right respectively. In real analysis, a user would not have prior knowledge of the exact nature of the source spectrum. Hence, we fit two models to the data in each case: model \texttt{m1}, the ``pure Band'' model (``\texttt{grbm}''), and model \texttt{m2} (``\texttt{grbm+bb}''), which incorrectly assumes the presence of a blackbody component. The second row shows residuals (blue) after fitting the spectrum with model \texttt{m1}, while the third row shows residuals (red) after fitting the spectrum with model \texttt{m2}. The bottom row (green) is the residuals corresponding to the \texttt{grbm} component of  \texttt{m2}, obtained after fitting the spectrum with the full model \texttt{m2}, then deleting the \texttt{bb} part (similar to plots used to show line features in X-ray analyses).

\begin{table*}[!htbp]
    \def\arraystretch{1.5}
    \centering
    \begin{tabular}{|cc|c|c|c|c|c|c|}
    \hline
    \hline
    Model & Parameter & \multicolumn{2}{c|}{100\% \texttt{ARM}} & \multicolumn{2}{c|}{90\% \texttt{ARM}} & \multicolumn{2}{c|}{110\% \texttt{ARM}}\\
    \hline
     & &  \multicolumn{6}{c|}{Best fit values}\\
    \hline
     & & \texttt{m1} & \texttt{m2} & \texttt{m1} & \texttt{m2} & \texttt{m1} & \texttt{m2}\\
    \hline
    \multirow{4}{*}{Band} & alpha & -0.5 $\pm$ 0.01 & -0.52 $\pm$ 0.02 & -0.54 $\pm$ 0.12 & -0.52 $\pm$ 0.02 & -0.47 $\pm$ 0.01 & -0.53 $\pm$ 0.04\\
    & beta & -2.27 $\pm$ 0.04 & -2.32 $\pm$ 0.05 & -2.27 $\pm$ 0.04 & -2.32 $\pm$ 0.05 & -2.28 $\pm$ 0.04 & -2.33 $\pm$ 0.06\\
    & $E_{c} (keV)$ & 542 $\pm$ 12  & 575 $\pm$ 28& 553 $\pm$ 12.5 & 572 $\pm$ 22 & 531 $\pm$ 11& 601 $\pm$ 48\\
    & norm/$10^{-3}$ & 42.5 $\pm$ 0.25 & 40.3 $\pm$ 0.9 & 44.1 $\pm$ 0.25 & 41.0 $\pm$ 0.8 & 40.9 $\pm$ 0.24& 38.2 $\pm$ 1.35\\
     \hline
     \multirow{2}{*}{Black} & $kT (keV)$ & -- & 67.2 $\pm$ 15.1 & -- & 52.3 $\pm$ 7.8 & -- & 94.3 $\pm$ 13.7\\
    Body & norm & -- & 1.3 $\pm$ 0.7 & -- & 1.43 $\pm$ 0.4 & -- & 2.5 $\pm$ 1.6\\
     \hline
    \multicolumn{2}{|c|}{ \rough{$\chi^2$ (DOF)}} & \rough{560 (550)} & \rough{554 (548)} & \rough{566 (550)} & \rough{542 (548)} & \rough{560 (550)} & \rough{555 (548)}\\
     \hline
    \end{tabular}
    \caption{Best-fit values for all the parameters with their 1-$\sigma$ errors and the reduced $\chi^2$ of the fits as described in \S\ref{sec:pureband} (pure band incidence) for all the three cases mentioned in the \S\ref{sec:inexact_ARM}. The two models considered here are \texttt{grbm} and \texttt{bb}. Model m1 corresponds to a pure \texttt{grbm} spectrum while model m2 reflects an additive \texttt{bb} component in addition to the \texttt{grbm} (consistent with the Figure~\ref{fig:band_sc}). Description of the parameters: \texttt{$\alpha$} and \texttt{$\beta$} are rising and decaying spectral indices, $E_c$ is the characteristic energy; related to peak energy $E_p$ by $E_p$ = $E_c$ (2 + $\alpha$). \textit{T} represents the temperature of the black body where \textit{k} is the Boltzmann constant. Both $E_c$ and \textit{kT} are expressed in keV. \textit{norm} is the spectral norm for each of the components of the model and is expressed in $\mathrm{photons~cm^{-2}~s^{-1}~keV^{-1}}$.}
    \label{tab:band_sc_fit_param}
\end{table*}

It can be seen from Table~\ref{tab:band_sc_fit_param} (and Figure~\ref{fig:band_sc}) that when 100\% \texttt{ARM} is considered, we accurately recover the incident band parameters. Further if we try to include a \texttt{bb} component in this case, the $\chi^2$ value decreases only marginally by 6 for 2 additional Degrees of Freedom (DOF) from model \texttt{m1} to model \texttt{m2}.
On the other hand, when we use an underestimated \texttt{ARM}, \rough{the best-fit (with \texttt{m1}) band component appears softer ($\alpha$ decreases) as compared to the incident pure band spectrum. Addition of a black body component (using model \texttt{m2}) improves the quality of the fit  with prominent blackbody component being ``detected'' in data, though the source model is a pure band spectrum (Figure~\ref{fig:band_sc}, middle column). In this case, the $\chi^2$ value decreases by 24 for two additional \texttt{DOF}s.} This additive model component is shown by the solid red line in the top panel, and also by the green ``\texttt{grbm residuals (m2)}'' sub-plot. The error estimate on the blackbody norm indicates a statistically significant fit --- arising out of ignored / unknown systematic errors in the \texttt{ARM}.
Lastly, when we use overestimated \texttt{ARM}, the best-fit (with \texttt{m1}) band component  appears harder ($\alpha$ increases) as compared to the injected source spectrum. If we directly compared this with the source spectrum, we would have seen negative residuals \rough{(not shown here)} as the \texttt{ARM} used in fitting over-predicts the observed spectrum. \rough{Adding a blackbody component to this should imply that we are trying to fit a negative one. Even then, if we try to force-fit such a model to the data, we get a fit with the band component becoming much softer and unrealistic realizations for the blackbody parameters are obtained (evident from the unusually large \textit{kT} value and large error in the norm of the balckbody in Table~\ref{tab:band_sc_fit_param})}.

%\rough{Adding a blackbody component to this produces a statistically insignificant fit (seen from the errors on the model parameters in Table \ref{tab:band_sc_fit_param}) since it implies that we are trying to fit a negative blackbody component. If we try to fit such a model to the data, the band component becomes softer and unrealistic realizations for the blackbody parameters are obtained. (evident from the \textit{kT} value of the balckbody and the errors on the model parameters in Table \ref{tab:band_sc_fit_param})}.

\subsubsection{Band and blackbody spectral model:}\label{sec:grbmbb}

We now consider a case where the GRB intrinsically has a blackbody component. We simulate a source spectrum with the same band function as before, and an additive blackbody characterised by \textit{kT}=40 keV and \textit{norm}=5. The temperature and relative intensity of this component are consistent with typical blackbody components detected in some GRBs.

\begin{figure*}[htp]
    \hspace{-2cm}\includegraphics[scale=0.7]{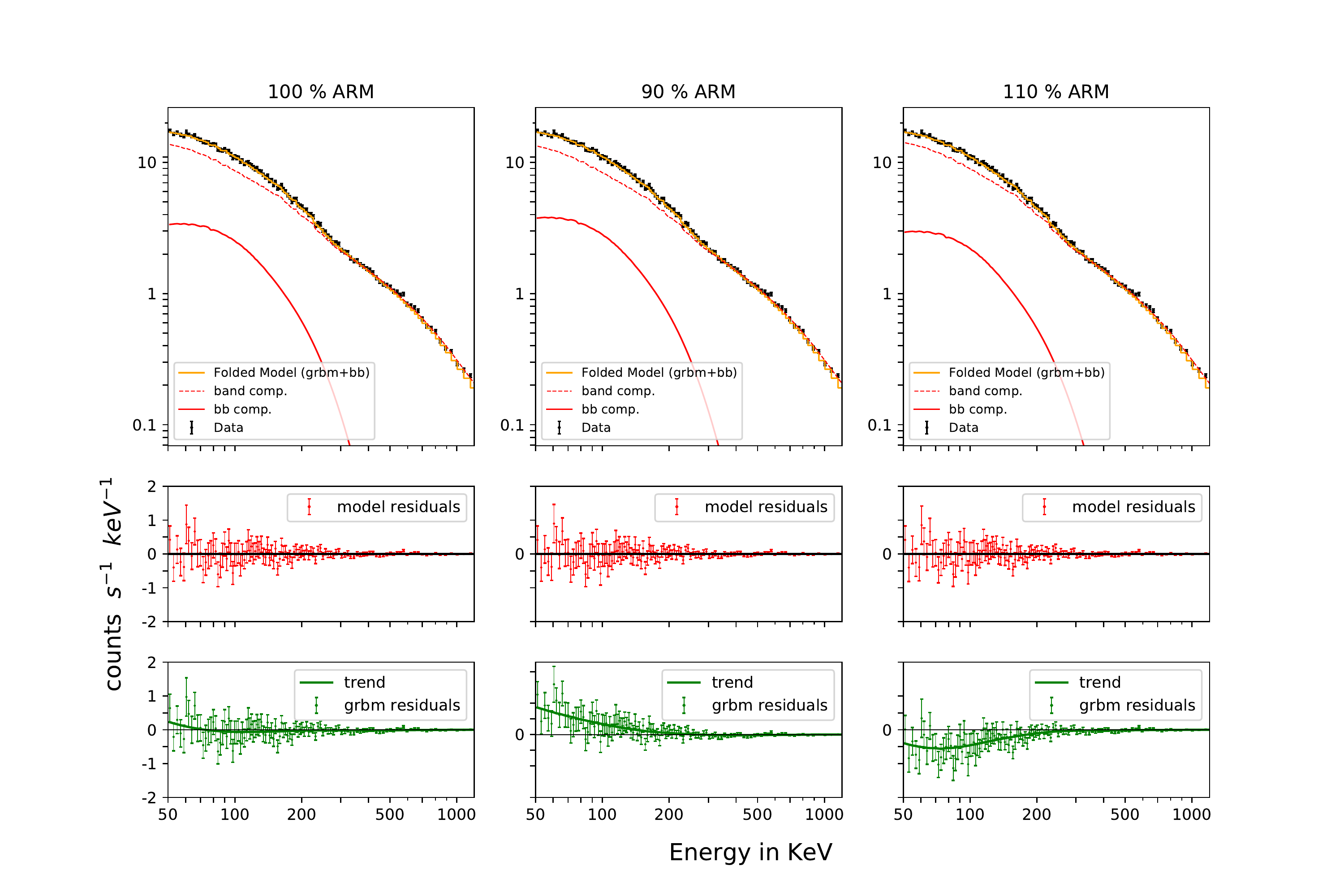}
    \caption{Simulations of the total GRB spectrum (incident+scattered) and the corresponding \xs\ fits to the simulated data for Band + Blackbody incidence. Each column represents a different response matrix (Atmospheric Response Matrix (\texttt{ARM}) convolved with the detector response matrix (\texttt{DRM})). The first column represents the accurate \texttt{ARM} and the next two columns correspond to \texttt{ARM} which is under and over estimated by 10\%. The incident spectrum is a short GRB modelled as a band function with parameters $\alpha = -0.5$, $\beta = -2.25$, and $E_\mathrm{peak} = 800$~keV ($E_c=533$~keV) and \textit{norm} = 0.0427 and an additive black body component with \textit{kT} = 40~keV and \textit{norm} = 5. The simulated data are marked in \rough{black} (with error bars overlaid). The solid orange line denotes the \xs\ fit for the simulated data. The data is modelled by a band function (\texttt{grbm}) with an additive blackbody (\texttt{bb}) component. The top row shows the \xs\  fits to the simulated data with the fitting model (\texttt{grbm + bb}) shown in red. The second row (from top) shows the residual plots (red) when \texttt{grbm + bb} model is used to fit the data. \rough{Although all the residuals look similar they do exhibit differences and the differences are small compared to the value of the residuals themselves. The bottom row shows the residuals (green) of the additional \texttt{grbm} component of the model. These are obtained by fitting the model to the data and then deleting the incident spectrum, thus highlighting the presence of an additional residual black body (if any), with the additional contribution from the black body component itself overlaid marked as \textit{`trend'}.} The best-fit parameter values are given in Table~\ref{tab:band_bb_sc_fit_param}.
}
    \label{fig:band_bb_sc}
\end{figure*}

Since the source has an intrinsic blackbody, fits with a pure band function always give poor results and we do not disucuss those. Instead, we focus our attention on modelling the spectrum with the \texttt{m2} model comprising of the band spectrum and a blackbody (Figure~\ref{fig:band_bb_sc}, Table~\ref{tab:band_bb_sc_fit_param}). As before, we find that if the \texttt{ARM} is correct, then the parameters are recovered well. However, any errors in the \texttt{ARM} will alter the values: for instance, if the \texttt{ARM} is underestimated, we find a brighter blackbody (see Table~\ref{tab:band_bb_sc_fit_param} and bottom row of the column at the middle of Figure~\ref{fig:band_bb_sc}), and conversely if the \texttt{ARM} used in the analysis overestimates the Earth reflection component, the inferred blackbody is fainter than the real value (shown by the plot at bottom-right of Figure~\ref{fig:band_bb_sc}).  The best fit values for the model parameters are shown in Table~\ref{tab:band_bb_sc_fit_param}. It can be seen from Table~\ref{tab:band_bb_sc_fit_param} that the best-fit values of the band spectrum are apparently the same in all the cases, with the error in \texttt{ARM} reflected mostly in the \texttt{bb} component (\rough{can be seen from the \textit{norm} of the blackbody component}).

\begin{table*}[!htbp]
    \def\arraystretch{1.5}
    \centering
    \begin{tabular}{|cc|c|c|c|c|c|c|}
    \hline
    \hline
    Model & Parameter &  & 100\% \texttt{ARM} & 90\% \texttt{ARM}& 110\% \texttt{ARM}\\
    \hline
     & & Incident & \multicolumn{3}{c|}{Best fit values}\\
    \hline
    \multirow{4}{*}{Band} & alpha & -0.5 & -0.54 $\pm$ 0.04 & -0.55 $\pm$ 0.04 & -0.54 $\pm$ 0.04\\
    & beta & -2.25 & -2.26 $\pm$ 0.04 & -2.26 $\pm$ 0.05 & -2.26 $\pm$ 0.04\\
    & $E_{c} (keV)$ & 533 & 562 $\pm$ 23 & 567 $\pm$ 24 & 557 $\pm$ 23\\
    & norm/$10^{-3}$ & 42.7 & 43.4 $\pm$ 1.1 & 43.6 $\pm$ 1.0 & 43.3 $\pm$ 1.0\\
     \hline
     \multirow{2}{*}{Black Body} & $kT (keV)$ & 40 & 40.8 $\pm$ 1.5 & 40.2 $\pm$ 1.3 & 41.4 $\pm$ 1.8\\
     & norm & 5 & 4.7 $\pm$ 0.4 & 5.31 $\pm$ 0.37 & 4.08 $\pm$ 0.37\\
     \hline
    \multicolumn{2}{|c|}{\rough{$\chi^2$ (DOF)}} & -- & \rough{528 (548)} & \rough{530 (548)} & \rough{528 (548)}\\
    \hline
    \end{tabular}
    \caption{Best-fit values for all the parameters with their 1-$\sigma$ errors and the reduced $\chi^2$ of the fit as described in \S\ref{sec:grbmbb} (Band and \texttt{bb} incidence) for all the three cases mentioned in the \S\ref{sec:spec_grb}. The model considered here is \texttt{grbm} with an additive \texttt{bb} (consistent with the Fig. \ref{fig:band_bb_sc}). Description of the parameters: \texttt{$\alpha$} and \texttt{$\beta$} are rising and decaying spectral indices, $E_c$ is the characteristic energy; related to peak energy $E_p$ by $E_p$ = $E_c$ (2 + $\alpha$) of the Band function. \textit{T} represents the temperature of the black body where \textit{k} is the Boltzmann constant. \textit{norm} is the spectral norm for each of the components of the model and is expressed in $\mathrm{photons~cm^{-2}~s^{-1}~keV^{-1}}$.}
    \label{tab:band_bb_sc_fit_param}
\end{table*}

\section{Conclusion and future work}\label{sec:discussions}

In this study, we have presented  some of our simulation outcomes helpful in 
better understanding the nature of the Earth's atmospheric reflection 
of X-ray/\gray\ photons from transient space objects and how can it affect the 
observation. We are in the process of developing a comprehensive 
and open-source simulation framework for estimating the atmospheric scattered 
component to be used in the construction of a publicly available database on 
Atmospheric Response Matrix (ARM). This particular study concentrates on 
contemplating the significance of proper evaluation of the atmospheric 
reflection component in terms of its effect on prompt GRB spectral analysis.
In some upcoming studies we intend to perform simulations to find out the influence 
of Earth's atmospheric scattering on the determination of polarization of 
high energy photons from such sources.

We find that the Earth's atmosphere is quite effective at scattering incident
X-rays and \gray s, creating an overall reflected spectrum that is usually strong 
in the 30--300~keV region.  The GRB fireball model \citep{Paczynski_1986, 
Goodman_1986, Lundman_2013}  predicts thermal components in GRB spectra. 
Such components have been identified in a multitude of GRBs 
\citep[for instance][]{Ryde_2005, Guiriec_2011, Guiriec_2013, Iyyani_2013, 
Iyyani_2015, Axelsson_2012, Burgess_2014, Nappo_2017}, and their temperatures 
are similar to the ones obtained in our fits above. The simplified example 
in \S\ref{sec:spec_grb} demonstrates how Earth's reflection can mimic such 
a thermal component. While this component is typically modelled in 
analysis, \S\ref{sec:inexact_ARM} shows how even small systematic errors 
in the estimation of atmospheric reflection can alter our interpretations of GRB data.

The problem may be further exacerbated when accounting for the direction-dependent 
sensitivity of the detector instrument. For instance, the effective area of \asat-CZTI 
varies by more than two orders of magnitude over the entire sky \citep{Mate2021}. If the GRB were to be incident from a low-sensitivity direction while 
the scattered radiation arrives from a high-sensitivity direction for the satellite, 
the scattered component can be an order of magnitude larger than the incident 
component --- further magnifying the effect of any uncertainties in the scattering model.

For typical observations for which one needs smaller contribution of thermal
(blackbody) flux to account for the hump in the intermediate range of prompt
GRB spectra, extra care should be taken to exclude any contribution from
atmospheric scattering. For example, for GRB110721A
\citep{Axelsson_2012,Iyyani_2013}, blackbody is found to contribute maximum of
$\sim$ 10\% of total flux, for GRB1000724B \citep{Guiriec_2011} this is found
to be $\sim$4\%. In these cases a slight over/under estimation of the
scattering contribution may impose considerable artefacts in the analysis
leading to wrong estimation of the thermal component.

This underscores the need for having robust \texttt{ARM} calculations 
that can be used in any spectral analysis. Likewise, such calculations are also 
needed for localisation and polarisation analyses. The calculation of the reflected 
spectrum from the Earth involves several complexities like creating an accurate 
description of the atmosphere, accounting for all physical effects, and generating 
a large number of incident photons to get good statistics. Approximations have to 
be made at each stage to make the problem tractable, and each approximation may 
introduce systematic errors in the final answer, which can have significant impacts 
on the scientific outcomes.

It is not feasible for each mission to dedicate significant resources and 
undertake simulations to calculate the \texttt{ARM}, nor should it be necessary. 
Instead, it is important that a few groups independently develop response matrices, 
and compare their results to quantify accuracy and reliability of the answers. 
Further, end users should be made aware of these limitations, and inclusion of 
appropriate systematic errors in analysis may be recommended.

 Going further, there are several improvements we intend to undertake. Firstly, 
there is a direct trade-off between using a large detector for collecting 
sufficient number of photons, and the resultant large differences in the scattering 
angles of photons that are incident on opposite ends of the detector. We will explore 
more simulation geometries, including multiple-detector scenarios, that will allow us 
to more effectively utilise all data generated in a simulation.

Second, the current simulations were carried out over a relatively coarse
energy grid, at a certain incidence angle, with a modest number of photons. We
plan to undertake convergence studies to determine the step sizes required in
energies and angles to ensure reliable simulation results. We will also quantify
the relationship between the number of photons in each monochromatic beam and
the statistical uncertainty in the final reflected spectrum. Once these
requirements are clearly defined, we will undertake simulations to create a
database that can be re-used for all further calculations. 

The final source codes and outputs of our simulations will be made available 
publicly for scrutiny and reuse, in a format that may be easily repurposed for other missions. 

\appendix

\section*{Acknowledgements}
CZT--Imager is built by a consortium of Institutes across India. The Tata Institute of Fundamental Research, Mumbai, led the effort with instrument design and development. Vikram Sarabhai Space Centre, Thiruvananthapuram provided the electronic design, assembly and testing. ISRO Satellite Centre (ISAC), Bengaluru provided the mechanical design, quality consultation and project management. The Inter University Centre for Astronomy and Astrophysics (IUCAA), Pune did the Coded Mask design, instrument calibration, and Payload Operation Centre. Space Application Centre (SAC) at Ahmedabad provided the analysis software. Physical Research Laboratory (PRL) Ahmedabad, provided the polarisation detection algorithm and ground calibration. A vast number of industries participated in the fabrication and the University sector pitched in by participating in the test and evaluation of the payload.

The Indian Space Research Organisation funded, managed and facilitated the project.

Sourav Palit wants to thank Indian Institute of Technology Bombay (IITB) for providing the scholarship 
and necessary resources to be able to perform this study.

%\clearpage
%%References section
%\begin{theunbibliography}{}
\bibliography{reference}
\balance
%\vspace{-1.5em}

%\end{theunbibliography}

\end{document}